\documentclass[11pt]{article}
\usepackage{amssymb,amsthm,amsmath}
\usepackage{xcolor,paralist,hyperref,titlesec,fancyhdr,etoolbox}
\usepackage{graphicx}
\usepackage{booktabs}
\usepackage{multirow} 
\usepackage{longtable}
\usepackage{indentfirst,csquotes}
\usepackage{lscape}

\usepackage{fancyhdr}
\begin{document}

% Full title of the paper (Capitalized)
\title{Muonic hyperfine structure and the Bohr-Weisskopf effect}

  \author{J.R. Persson\\
  Department of Physics\\
  NTNU\\
  NO-7491 Trondheim\\
  Norway\\
  E-mail: jonas.persson@ntnu.no}
\date{24.10.2025}
 \maketitle 

% Abstract (Do not insert blank lines, i.e. \\) 
\begin{abstract}
    An update is given on the experimental values of the magnetic hyperfine structure and the Bohr-Weisskopf effect in muonic atoms. The need for more measurements and systematic calculations is discussed to allow the differentiation of different models of the Bohr-Weisskopf effect in nuclei.
\end{abstract}

\newpage

\section{Introduction}
Muons have since their discovery in 1936 been extensively studied and have been used in the study of nuclear properties\cite{knecht2020study} and different applications, for example, muon spin spectroscopy \cite{blundell2021muon} and muon tomography \cite{Alvarez1970SearchFH}. Since the muon is in many aspects similar to the electron, with the same charge but $\approx 207$ times heavier, this means that when a muon comes close to an atom it might be captured and form a hydrogen-like muonic atom. The lifetime of the muon ($2.2\mu s$) is long enough for being able to cascade down to the lowest orbit and can be considered a stable particle in the atomic timescale. This will give rise to large effects on the energy levels due to the extension of the nucleus,  i.e. the distribution of charge and magnetisation, especially for heavier nuclei, as well as relativistic effects.\\
As a muonic atom can be considered to be hydrogen-like it is possible to do high-accuracy calculations of the level structure which in combination with experimental results can give information on different nuclear parameters like nuclear charge radii \cite{engfer1974charge,piller1990nuclear}, quadrupole moments \cite{tanaka1982ground} and magnetic dipole interaction constants \cite{ruetschi1984magnetic}.\\
 In this paper, we focus on the use of muons to probe the nucleus and more specifically the distribution of magnetisation. The finite-nucleus effect of magnetisation or the Bohr-Weisskopf (BW) effect \cite{Bohr1951,Weisskopf1950} is essential in the determination of accurate values of the nuclear magnetic dipole moment if it cannot be determined directly but rather through the hyperfine structure (hfs) as is the case in short-lived isotopes \cite{Persson2013}. Knowledge of the distribution of magnetisation is also important in studies of QED effects as well as studies of atomic parity violation \cite{sanamyan2023empirical}, where there is a need for high-quality modelling of the atomic wave functions within the nucleus. This means that the nuclear models used must take both the charge and magnetisation distribution into account.\\
 Since the magnetic hyperfine interaction splitting in the $2p_{1/2}-1s_{1/2}$ transitions in heavy muonic atoms are in the same order of magnitude as the energy resolution in the detectors used, the splitting has to be determined from the line broadening. This line shape analysis is complicated and requires the use of simultaneously recorded calibration lines. The magnetic hyperfine splitting in transitions between higher-lying states is usually too small to be detected, why the $2p_{1/2}-1s_{1/2}$ transitions are the only option.\\
 Due to the experimental difficulties, there does not exist many measurements of the magnetic hyperfine structure and the majority were done during a rather short period during the 1960s and 1970s. The interest in measuring the nuclear charge radii continued and indicated that measurements of the magnetic hyperfine structure are possible, with the muX project \cite{wauters2021mux} as an example. This paper aims to give an update of the previously published reviews in 1974 \cite{engfer1974charge} and 1984\cite{Buttgenbach1984} and hopefully initiate new measurements and more interest in the topic. Even if the number of new measurements is limited it is justified with an update.\\

 The published experimental $A$ constants are presented in table \ref{Measured A constants}. It should be noted that the $A$ constants in previous reviews \cite{engfer1974charge, Buttgenbach1984} were defined in different ways. In this paper we follow the normal definition of the $A$ constant for muonic $s$ and $p_{1/2}$ states, also used in \cite{Buttgenbach1984}:
     \begin{equation}
    A=\frac{\Delta E}{F'}
    \end{equation}
with $\Delta E$ being the magnetic hfs splitting and $F'=I+\frac{1}{2}$, where $I$ being the nuclear spin. The definition used in \cite{engfer1974charge} is:
\begin{equation}
    A=\frac{\Delta E}{\frac{2I+1}{I}}
    \end{equation}
Something that might cause some confusion.\\
Even if the number of measurements is quite small, they might still give valuable information on the quality of the different nuclear models used to calculate the BW-effect. However, we will argue that there is a need for more measurements, but also systematic calculations with different nuclear models to expand calculations to atomic systems\cite{jonsson2022introduction}.

\begin{table}[h!p]
\centering
\caption{Experimental and recommended values of the A constant in the muonic $1s_{1/2}$ states. Bold indicates updated value compared to ref.\cite{Buttgenbach1984}.\label{Measured A constants}}
%	\begin{adjustwidth}{-\extralength}{0cm}
		
		\begin{tabular}{c c c c c}

			\textbf{Isotope}	& \textbf{I}	& \textbf{A constant}     & \textbf{reference} & \textbf{Recommended A }\\
   		
			& & \textbf{(keV)}     &  & \textbf{constant (keV)}\\
			\midrule
\multirow[m]{2}{*}{$^{93}$Nb}	& \multirow[m]{2}{*}{9/2}	& 0.716(49)		& \cite{cheng1971magnetic}\\
			  	         &          			   & 0.693(22)	   & \cite{povel1973m1} & 0.697(20)\\
\midrule
\multirow[m]{2}{*}{$^{115}$In} & \multirow[m]{2}{*}{9/2}	& 0.733(32)		& \cite{lee1969finite}\\
			  	         &          			   & 0.716(14)	   & \cite{link1971magnetic,link1974magnetic} & \textbf{0.719(13)}\\
\midrule
\multirow[m]{1}{*}{$^{127}$I} & \multirow[m]{1}{*}{5/2}	& 0.712(72)		& \cite{lee1971resonance}& 0.712(72)	\\
\midrule                   
\multirow[m]{1}{*}{$^{133}$Cs} & \multirow[m]{1}{*}{7/2}	& 0.634(103)		& \cite{lee1969finite}& \textbf{0.634(103)}\\
\midrule  
\multirow[m]{2}{*}{$^{139}$La}	& \multirow[m]{2}{*}{7/2}	& 0.691(52)		& \cite{thompson1969electromagnetic}\\
			  	         &          			   & 0.771(46)	   & \cite{cheng1971magnetic}& 0.736(40)\\
\midrule
\multirow[m]{2}{*}{$^{141}$Pr}	& \multirow[m]{2}{*}{5/2}	& 1.22(13)		&\cite{lee1969finite} \\
			  	         &          			   & 1.216(48)	   & \cite{johnson1969magnetic}& 1.216(45)	\\
\midrule
\multirow[m]{1}{*}{$^{151}$Eu} & \multirow[m]{1}{*}{5/2}	& 0.64(22)		& \cite{carrigan1968muonic2}& 0.64(22)\\
\midrule 
\multirow[m]{1}{*}{$^{190}$Os} & \multirow[m]{1}{*}{2}	& 0.266(32)		& \cite{link1972magnetic}& 0.266(32)\\
\midrule 
\multirow[m]{1}{*}{$^{192}$Os} & \multirow[m]{1}{*}{2}	& 0.320(32)		& \cite{link1974magnetic}& 0.320(32)	\\
\midrule 

\multirow[m]{1}{*}{$^{191}$Ir} & \multirow[m]{1}{*}{3/2}	& 0.44(8)		&\cite{link1974magnetic}& 0.440(80) \\
\midrule 
\multirow[m]{1}{*}{$^{191m}$Ir} & \multirow[m]{1}{*}{5/2}	& 0.584(64)		& \cite{link1974magnetic}& 0.584(64)\\
\midrule 

\multirow[m]{1}{*}{$^{193}$Ir} & \multirow[m]{1}{*}{3/2}	& 0.307(54)		& \cite{link1974magnetic}&\textbf{ 0.307(54)}\\
\midrule 
\multirow[m]{1}{*}{$^{193m}$Ir} & \multirow[m]{1}{*}{5/2}	& 0.416(48)		& \cite{link1974magnetic}& \textbf{0.416(48)}\\
\midrule
\multirow[m]{1}{*}{$^{197}$Au} & \multirow[m]{1}{*}{3/2}	& 0.23(4)	&\cite{POWERS1974413}& \textbf{0.23(4)}	 \\
\midrule 
\multirow[m]{1}{*}{$^{199}$Hg} & \multirow[m]{1}{*}{1/2}	& 0.47(12)		&\cite{link1974magnetic}& 0.47(12)	\\
\midrule 
\multirow[m]{1}{*}{$^{200}$Hg} & \multirow[m]{1}{*}{2}	& 0.262(42)		& \cite{link1974magnetic}& \textbf{0.262(42)}\\
\midrule 
\multirow[m]{1}{*}{$^{203}$Tl} & \multirow[m]{1}{*}{1/2}	& 2.340(80)		&\cite{backe1974nuclear}& 2.340(80)	 \\
\midrule 
\multirow[m]{2}{*}{$^{205}$Tl}	& \multirow[m]{2}{*}{1/2}	& 2.300(32)		& \cite{backe1972study}\\
			  	         &          			   & 2.44(12)	   & \cite{cheng1971magnetic}& 2.309(35)	\\
\midrule
\multirow[m]{4}{*}{$^{209}$Bi}	& \multirow[m]{4}{*}{9/2}	& 1.11(23)		& \cite{powers1968hyperfine}\\
			  	         &          			   & 0.933(90)	   & \cite{carrigan1968hyperfine1}\\
			  	         &          			   & 0.96(7)	   & \cite{lee1972nuclear}& \textbf{	0.905(27)}\\
			  	         &          			   & 0.888(30)	   & \cite{ruetschi1984magnetic}\\
\midrule
\multirow[m]{1}{*}{$^{209m}$Bi} & \multirow[m]{1}{*}{13/2}	& 0.414(59)		& \cite{ruetschi1984magnetic}& \textbf{0.414(59)}\\
\midrule 
\multirow[m]{1}{*}{$^{209m}$Bi} & \multirow[m]{1}{*}{9/2}	& 0.420(77)		&\cite{ruetschi1984magnetic}& \textbf{0.420(77)	}\\
\midrule 
\multirow[m]{1}{*}{$^{209m}$Bi} & \multirow[m]{1}{*}{15/2}	& 0.300(70)		&\cite{ruetschi1984magnetic}& \textbf{0.300(70)	}\\
\midrule 

			\bottomrule
		\end{tabular}

\end{table}

%%%%%%%%%%%%%%%%%%%%%%%%%%%%%%%%%%%%%%%%%%
\section{Hyperfine structure in muonic atoms}
The hyperfine structure in muonic atoms is due to the electric quadruple (E2) and magnetic dipole (M1) interactions. The electric quadrupole moment depends on the effective shape of the nucleus while the magnetic dipole interaction depends on the magnetic dipole moment of the nucleus. Since the magnetic dipole moment of the muon is $\approx$ 207 times smaller than the magnetic moment of the electron, the magnetic dipole interaction is normally two orders of magnitude smaller than the quadrupole interaction. This implies that the magnetic part may only be studied in cases where the total angular momenta are 1/2 for either I or J in one of the states in a transition, making the transitions $2p_{1/2,3/2} \rightarrow 1s_{1/2}$ and the $1s_{1/2}$ state best suited for studies. Considering that the transition energies are in the MeV region and the A constants are in the keV region, the experimental difficulties are quite challenging. This can in some respect explain why the technique fell out of fashion, especially with the possibility to do measurements on hydrogen-like highly charged ions. The difficulty is also evident from the errors in table \ref{Measured A constants}. Based on the available data we have done an assessment based on the quality of the individual data and taken the weighted arithmetic average as the recommended values for the experimental $A$ constants in table \ref{Measured A constants}

\section{Hyperfine anomaly in muonic atoms}

For the low-lying states, the point-like nucleus is not a good approximation and the finite size must be taken into account in calculations of the hyperfine structure. This means the relativistic Dirac equation must be solved with a reasonable nuclear potential. The effect of an extended charge distribution is normally referred to as the Breit-Rosenthal (BR) effect \cite{Rosenthal1932} and is calculated with different models. The most common charge distribution used is the two-parameter Fermi distribution given as
  \begin{equation}
    \rho (r) = \rho_0 \left\{1+exp (4\; ln 3 \: \frac{r-c}{t})\right\}^{-1}
    \end{equation}
where $t$ is the skin thickness and $c$ is the half-density radius of the nuclei. This distribution is normally not a good approximation for deformed nuclei where a multipole expansion can be used \cite{GUSTAVSSON1998343}. 
The distribution of magnetisation or BW-effect is normally extracted from experimental data and calculations of the A constant for a distributed charge (including the BR-effect) and a point-like dipole ($A_{0,p.d.}$). The experimental A constant is thus described as
\begin{equation}
\label{a factor}
    A_{exp} =A_0 (1+\epsilon_{BR}) (1+\epsilon_{BW}) = A_{0,p.d.}(1+\epsilon_{BW})
    \end{equation}
where the BW-effect should be independent of the charge distribution used for the BR-effect. There might be a dependence, but that is not significant in comparison with other sources of uncertainty.  However, when extracting the BW-effect from experimental data, the values will depend on the calculations why it is important to have both a reasonable charge distribution and high-quality calculations. To obtain the experimental BW-effect ($\epsilon_{BW, exp}$) one divides the experimental $A$ constant with the calculated point dipole A constant ($A_{0,p.d.}$).
\begin{equation}
   \epsilon_{BW,exp}=\frac{A_{exp}}{A_{0,p.d.}} -1
    \end{equation}
As this value depends on the calculation it is important that this is done with high precision and that the magnetic dipole moment, charge distribution and parameters used are presented in detail.
In table~\ref{epsilon}  the experimental BW-effect is presented with different calculations of the point dipole A constant and the values of the magnetic dipole moment used. As the calculated point magnetic hyperfine A constant ($A_{0,p.d.}$) differs in the original papers one obtains slightly different values of the BW-effect.

%%%%%%%%%%%%%%%%%%%%%%%%%%%%%%%%%%%%%%%%%%
\section{Calculations of the BW-effect}

There exist several different models for calculating the BW-effect, see \cite{Buttgenbach1984} for an overview. Muonic atoms are more sensitive to the distribution of magnetisation and relatively easy to do calculations on, why muonic atoms are expected to be better to use in comparing different models. In the case of muonic atoms, only a few different models have been used systematically in cases where there also exist experimental values of the magnetic dipole interaction. \\
It is therefore possible to check how good the different models are at reproducing the BW-effect. Systematic calculations have been done by Johnson and Sorensen \cite{johnson1969magnetic} using an expansion of the single-particle model, the configuration mixing model (CM)\cite{blin1953magnetic, arima1954configuration} and two extensions with pairing (PM and PQM). Fujita and Arima \cite{fujita1975magnetic} applied their microscopic model to two different sets of nuclear parameters (MT I and II) in the woods-Saxon potential used in their calculation.\\
Recently, Sanamyan et al. \cite{sanamyan2023empirical} applied both a uniform magnetisation distribution and single-particle models to the $^{133}$Cs,$^{203}$Tl,$^{205}$Tl and $^{209}$Bi muonic atoms to find the BW contribution to the A constant. Similar calculations were also done by Elizarov et al. \cite{elizarov2005hyperfine} on $^{203}$Tl$,^{205}$Tl and $^{209}$Bi muonic atoms. As the calculations of Sanamyan et al.\cite{sanamyan2023empirical} and Elizarov et al. \cite{elizarov2005hyperfine} are the most recent, these isotopes are presented separately in table \ref{epsilon selected} from the isotopes with only older calculations that are presented in table \ref{epsilon}. It should be mentioned that Oreshkina and collaborated have done calculations on the hyperfine structure in heavy muonic atoms for which there does not exist any experimental data\cite{michel2019higher,michel2017theoretical}.\\

It should be noted that the BW-effect has been obtained with the calculated point dipole $A_0$ constant given in the original paper. As these values differ both should be considered in the comparison. As the experimental errors are quite large most calculated values with the models are within the error bars. As can be seen in table \ref{epsilon} there are some exceptions where the models fail to reproduce the BW-effect. In $Eu$ the nucleus is deformed which might explain the deviation, this could also explain the deviation in Ir. \\
One might observe that it seems like the lower values of the nuclear spin yield larger differences between the experimental and calculated BW-effect than high spin nuclei, with the exception of Bi which is close to a magical number where the single-particle model should be valid. This might indicate that the distribution of magnetisation is harder to model than for high-spin nuclei. This could be because the wavefunction of the "high-spin" nucleon in the single particle model has a radial distribution with a maximum further from the origin and is easier to model in a proper way.\\
When considering the recent single-particle calculations with older calculations and experiments in table \ref{epsilon selected} we observe that the agreement is still good and on par with the best older calculations. One should also consider the homogeneous magnetisation model, which can be used to estimate the BW-effect. In the case of Tl (I=1/2), the single-particle and homogeneous magnetisation models are equally good. But since the homogeneous magnetisation (ball) model fails for Cs and Bi it is clear that this model is not appropriate for calculations of the BW-effect, this has also been observed in atoms \cite{roberts2021hyperfine}. However, it would be interesting to see how well a shell magnetisation can reproduce the BW-effect in muonic atoms and maybe replace the homogeneous magnetisation model as a simplistic model to use.
The single-particle model seems to reproduce the BW-effect reasonably well, with some minor issues for low-spin heavy nuclei, therefore, it would be interesting to perform systematic calculations for all measured nuclei to assess the model in detail.  \\
It is also possible to conclude that the calculated point dipole $A_0$ constant differs in the published papers, why it might also be important to perform systematic calculations with up-to-date programs. It is also clear that the simple model of a uniform magnetisation distribution is too crude to be used for muonic atoms, possible with a few exceptions. As the single-particle model seems to be the best to describe the BW-effect systematic calculations in muonic atoms will also be valuable.

\begin{table}[h!p]

\caption{Experimental and calculated values of the Bohr-Weisskopf effect $\epsilon$ in the muonic $1s_{1/2}$ states.
CM: configuration mixing model; PM: Pairing model; PQM: Pairing plus quadrupole; MT: Microscopic Model I and II.\label{epsilon}}
%	\begin{adjustwidth}{-\extralength}{0cm}
		
		\begin{tabular}{l|l|l|l|l|l|l|l|l|l|l}
			\toprule
			\textbf{Iso.}	& \textbf{I}	& \textbf{$\mu (\mu_N)$} & $A_0$& \textbf{$\epsilon_{exp}$} & \textbf{CM} & \textbf{PM} & \textbf{PQM} &  \textbf{MTI}  &  \textbf{MTII}    & \textbf{ref.}\\
			& &  &(keV) &  &  &  &  &   &   & \\   
			\midrule
\multirow[m]{2}{*}{$^{93}$Nb}	& \multirow[m]{2}{*}{9/2}	& \multirow[m]{2}{*} {6.1671} & 1.031 & -0.324(30) & -0.362 & -0.362 &  -0.332 & ~      & ~      & \cite{johnson1969magnetic} \\
			  	          &          			      &                             & 1.013 & -0.312(30) &~      & ~      &  ~      & -0.301 & -0.345 & \cite{fujita1975magnetic}\\
                \midrule

\multirow[m]{2}{*}{$^{115}$In} & \multirow[m]{2}{*}{9/2}	&{5.5351} & 1.156 & -0.378(18) & -0.392 & -0.392 &  -0.362 & ~      & ~      & \cite{johnson1969magnetic}  \\
			  	         &          			       & {5.5348} & 1.147 & -0.373(18) & ~      & ~     &  ~      & -0.335 & -0.384 & \cite{fujita1975magnetic}  \\
\midrule			  	        
 
\multirow[m]{2}{*}{$^{127}$I} & \multirow[m]{2}{*}{5/2}	&\multirow[m]{2}{*} {2.8091} & 1.120 &-0.364(101) & -0.329 & -0.336  & -0.207 & ~ & ~ & \cite{johnson1969magnetic}  \\
			  	         &          			       &                        &  1.145 &-0.378(101) & ~ & ~ & ~ & -0.333 & -0.368 &\cite{fujita1975magnetic}   \\
\midrule   
\multirow[m]{2}{*}{$^{133}$Cs} & \multirow[m]{2}{*}{7/2}	&\multirow[m]{2}{*} {2.5789} & 0.766 & -0.172(162) & -0.097 & -0.119 & -0.097 & ~ & ~ & \cite{johnson1969magnetic}  \\
			  	         &          			       &                            & 0.780 &-0.187(162)& ~ & ~ & ~ & -0.131 & -0.163 & \cite{fujita1975magnetic}   \\
\midrule
\multirow[m]{2}{*}{$^{139}$La} & \multirow[m]{2}{*}{7/2}	&\multirow[m]{2}{*} {2.7781} & 0.857 &-0.141(54) & -0.140 & -0.133 &  -0.133 & ~ & ~ & \cite{johnson1969magnetic} \\
			  	         &          			       &                            & 0.872 &-0.156(54)& ~ & ~ &  ~ & -0.154 & -0.189 & \cite{fujita1975magnetic}  \\
\midrule     
 \multirow[m]{2}{*}{$^{141}$Pr} & \multirow[m]{2}{*}{5/2}	& {4.28} &  1.928 &-0.369(37)& -0.390 & -0.390 &  -0.386 & ~ & ~ & \cite{johnson1969magnetic} \\
			  	         &          			       & {4.16} & 1.971 &-0.383(37)& ~ & ~ &  ~ & -0.366 & -0.403 &\cite{fujita1975magnetic} \\
\midrule         
  \multirow[m]{2}{*}{$^{151}$Eu} & \multirow[m]{2}{*}{5/2}	&\multirow[m]{2}{*} {3.463} & 1.632 &-0.608(334)& -0.397 & -0.392 &  -0.299 & ~ & ~ & \cite{johnson1969magnetic} \\
			  	         &          			       &                             & 1.707 &-0.625(334)& ~ & ~ &  ~ & -0.375 & -0.413 &\cite{fujita1975magnetic}  \\
\midrule

  {$^{190}$Os} &{2}	&{0.68} & 0.474 &-0.439(120)& ~ & ~ &  ~ & -0.370 & -0.426 &\cite{fujita1975magnetic}    \\
\midrule
 {$^{192}$Os} & {2}	& {0.78} &  0.567 &-0.436(100)& ~ & ~ &  ~ & -0.371 & -0.426 &\cite{fujita1975magnetic}   \\
\midrule 

 {$^{191}$Ir} &{3/2}	& {0.1454} &  0.140 &2.143(182)& ~ &  ~ & ~ & 2.910 & 3.058 &\cite{fujita1975magnetic}  \\
\midrule 
  {$^{191m}$Ir} &{5/2}	& {0.5} &  0.337 & 0.733(82)&~ & ~ &  ~ & -0.246 & -0.280 &\cite{fujita1975magnetic}    \\
\midrule 

 {$^{193}$Ir} & {3/2}	& {0.1637 } & 0.152 & 1.020(176)&~ &  ~ & ~ & 2.630 & 2.765 &\cite{fujita1975magnetic}   \\
\midrule 
 {$^{193m}$Ir} & {5/2}	& {0.6 } &   0.422 &-0.014(115)& ~ &  ~ & ~ & -0.286 & -0.320 &\cite{fujita1975magnetic}   \\
\midrule  
{$^{199}$Hg} & {1/2}	&{0.49787 } &  1.503 & -0.687(255)& ~ &  ~ & ~ & -0.811 & -0.838 &\cite{fujita1975magnetic}   \\
\midrule 
{$^{200}$Hg} & {2}	&{0.49?} &  0.321 &-0.184(160)& ~ & ~ & ~  & -0.302 & -0.338 &\cite{fujita1975magnetic}   \\
\midrule 

  \multirow[m]{2}{*}{$^{203}$Tl} & \multirow[m]{2}{*}{1/2}	& {1.61169} & 4.760 &-0.508(34)& -0.355 & -0.350 &  -0.347 & ~ &  & \cite{johnson1969magnetic} \\
			  	         &          			       &   {1.6115 } &  4.846 &-0.517(34)& ~ &  ~ & ~ & -0.419 & -0.451 &\cite{fujita1975magnetic}   \\
\midrule  
    \multirow[m]{2}{*}{$^{205}$Tl} & \multirow[m]{2}{*}{1/2}	&{1.62754} & 4.760 &-0.515(15)& -0.350 & -0.345   & -0.345 & ~ & ~ & \cite{johnson1969magnetic} \\
			  	         &          			       &  {1.6274 } &  4.868 &-0.526(15)& ~ & ~ &  ~ & -0.425 & -0.459 &\cite{fujita1975magnetic}  \\
\midrule  

    \multirow[m]{2}{*}{$^{209}$Bi} & \multirow[m]{2}{*}{9/2}	& {4.0794} & 1.351 &-0.330(30)& -0.322 & ~ & ~ & ~ & ~ & \cite{johnson1969magnetic}  \\
			  	         &          			       &   { 4.08 } &  1.385 &-0.347(30)& ~ &  ~ & ~ & -0.266 & -0.314 &\cite{fujita1975magnetic}   \\

			\bottomrule
		\end{tabular}
%	\end{adjustwidth}
%	\noindent{\footnotesize{* Tables may have a footer.}}
\end{table}
 \begin{landscape}

\setlength{\LTleft}{0pt}
\setlength{\LTright}{0pt}

\setlength{\tabcolsep}{0.5\tabcolsep}

\renewcommand{\arraystretch}{1.0}

\footnotesize % we need to squeeze the font size a lot!

\begin{longtable}{l|l|l|l|l|l|l|l|l|l|l|l}

\caption{Experimental and calculated values of the Bohr-Weisskopf effect $\epsilon$ in the muonic $1s_{1/2}$ states.
CM: configuration mixing model; PM: Pairing model; PQM: Pairing plus quadrupole; MT: Microscopic Model I and II; Ball: Homogeneous; SP: single-particle; SP-WS: Single-particle with Wood-Saxon.\label{epsilon selected}}\\

\textbf{Isotope} & \textbf{ $A_0$ (keV)} & {\textbf{$\epsilon_{exp}$} }& \textbf{CM}& \textbf{PM}& \textbf{PQM} &  \textbf{MTI}  &  \textbf{MTII} & \textbf{Ball} & \textbf{SP} & \textbf{SP-WS}   & \textbf{ref.}\\

\endhead

			\midrule
 
\multirow[m]{3}{*}{$^{133}$Cs}  & 0.766 & -0.172(162) & -0.097 & -0.119 & -0.097 & ~ & ~ & & & & \cite{johnson1969magnetic}  \\
			  	         &    0.780 &-0.187(162)& ~ & & & -0.131 & -0.163 & & & & \cite{fujita1975magnetic}   \\
                                & 0.762 & -0.172(162)&   & & &        &        & -0.413 &  -0.155 &-0.155 & \cite{sanamyan2023empirical}  \\          
\midrule

  \multirow[m]{3}{*}{$^{203}$Tl} &  4.760 &-0.508(34)& -0.355 & -0.350 &  -0.347 & ~ &  & & & & \cite{johnson1969magnetic} \\
			  	         &     4.846 &-0.517(34)& ~ &  ~ & ~ & -0.419 & -0.451 & & & &\cite{fujita1975magnetic}   \\
                                               & 4.712 & -0.503(34)&   & & &        &        & -0.527 &  -0.527 &-0.439 & \cite{sanamyan2023empirical}  \\ 
                                               & 4.70  & -0.502(34)&   & & &        &        &        &         & -0.425& \cite{elizarov2005hyperfine}\\
\midrule  
    \multirow[m]{3}{*}{$^{205}$Tl} & 4.760 &-0.515(15)& -0.350 & -0.345   & -0.345 & ~ & ~ & & & & \cite{johnson1969magnetic} \\
			  	         &      4.868 &-0.526(15)& ~ & ~ &  ~ & -0.425 & -0.459 & & & &\cite{fujita1975magnetic}  \\
                                                              & 4.744 & -0.513(15)&   & & &        &        & -0.527 &  -0.527 &-0.439 & \cite{sanamyan2023empirical}  \\ 
                                                             & 4.73  & -0.519(34)&   & & &        &        &        &         & -0.429& \cite{elizarov2005hyperfine}\\
\midrule  

    \multirow[m]{3}{*}{$^{209}$Bi} &  1.351 &-0.330(30)& -0.322 & ~ & ~ & ~ & ~ & & & & \cite{johnson1969magnetic}  \\
			  	         &       1.385 &-0.347(30)& ~ &  ~ & ~ & -0.266 & -0.314 & & & &\cite{fujita1975magnetic}   \\
                                   & 1.339 & -0.324(30)&   & & &        &        & -0.533 &  -0.310 &-0.347 & \cite{sanamyan2023empirical}  \\ 
                                   & 1.338  & -0.327(30)&   & & &        &        &        &         & -0.359& \cite{elizarov2005hyperfine}\\

			\bottomrule
%		\end{tabular}
	
%	\noindent{\footnotesize{* Tables may have a footer.}}
\end{longtable}
\end{landscape}

%%%%%%%%%%%%%%%%%%%%%%%%%%%%%%%%%%%%%%%%%%
\section{Discussion}
This update of measurements and theoretical calculations of the BW-effect in muonic atoms has unearthed the need for renewed systematic studies of the magnetic hyperfine interaction in the $1s_{1/2}$ states in muonic atoms. As the use of muonic atoms opens the possibility to obtain values of the BW-effect directly, new or improved measurements will make it possible to compare different nuclear models for the distribution of magnetisation. The muX project at PSI \cite{wauters2021mux} is such a project which might be able to do measurements in the lead and trans-lead regions. \\
In addition to the need for systematic measurements, there exists a need for high-precision calculations of the point dipole $A_0$ constant as illustrated by the difference in the published values presented in tables \ref{epsilon} and \ref{epsilon selected}, this will also give more certain values of the BW-effect ($\epsilon$). To find which of the nuclear models of magnetisation is best, the experimental values of the BW-effect should be compared with the values calculated using different models. This will possibly give a better comparison as one would expect that the muonic BW-effect is more sensitive to the distribution of magnetisation than electronic atoms. As of today, systematic calculations in muonic atoms only exist for the configuration mixing model \cite{johnson1969magnetic} and the microscopic model \cite{fujita1975magnetic} with a few exceptions \cite{sanamyan2023empirical, elizarov2005hyperfine}. It is important to do systematic calculations using other models to find which models are valid and for which isotopes. Doing so in muonic atoms will make it possible to choose the best model for calculations in atoms.

\bibliographystyle{plain}
\bibliography{ref}

\end{document}